# Discovery of homogeneously dispersed pentacoordinated Al$^V$ species on the surface of amorphous silica-alumina

Zichun Wang,[1] Yijiao Jiang,[2] Xianfeng Yi,[3] Cuifeng Zhou,[1] Aditya Rawal,[4] James Hook,[4] Zongwen Liu,[1] Feng Deng,[3] Anmin Zheng,[3,*] Alfons Baiker,[5] and Jun Huang[1,*]

[1] Laboratory for Catalysis Engineering, School of Chemical and Biomolecular Engineering, The University of Sydney, NSW 2006, Australia

[2] Department of Engineering, Macquarie University, Sydney, New South Wales 2109, Australia

[3] State Key Laboratory Magnetic Resonance and Atomic Molecular Physics, Wuhan Institute of Physics and Mathematics, Chinese Academy of Sciences

[4] Mark Wainwright Analytical Centre, University of New South Wales, Sydney, NSW 2052, Australia

[5] Institute for Chemical and Bioengineering, Department of Chemistry and Applied Bioscience, ETH Zürich, Hönggerberg, HCI, CH-8093 Zürich, Switzerland

*Abstract*

The dispersion and coordination of aluminium species on the surface of silica-alumina based materials are essential for controlling their catalytic activity and selectivity. $Al^{IV}$ and $Al^{VI}$ are two common coordinations of Al species in the silica network and alumina phase, respectively. $Al^V$ is rare in nature and was found hitherto only in the alumina phase or interfaces containing alumina, a behavior which negatively affects the dispersion, population, and accessibility of $Al^V$ species on the silica-alumina surface. This constraint has limited the development of silica-alumina based catalysts, particularly because $Al^V$ had been confirmed to act as a highly active center for acid reactions and single-atom catalysts. Here, we report the direct observation of high population of homogenously dispersed $Al^V$ species in amorphous silica-alumina in the absence of any bulk alumina phase, by high resolution TEM/EDX and high magnetic-field MAS NMR. Solid-state $^{27}Al$ multi-quantum MAS NMR experiments prove unambiguously that most of the $Al^V$ species formed independently from the alumina phase and are accessible on the surface for guest molecules. These species are mainly transferred to $Al^{VI}$ species with partial formation of $Al^{IV}$ species after adsorption of water. The NMR chemical shifts and their coordination transformation with and without water adsorption are matching that obtained in DFT calculations of the predicted clusters. The discovery presented in this study not only provides fundamental knowledge of the nature of aluminum coordination, but also paves the way for developing highly efficient catalysts.

Alumina and its mixed oxides are important catalytic materials both as active catalysts and as functional supports for active metal particles. The catalytic functions of these materials in chemical reactions are mainly dependent on the surface coordination of Al species due to their structure-activity relationship. Most research efforts have been focused on the tetrahedral and octahedral coordination ($Al^{IV}$ and $Al^{VI}$), which are the most popular coordinations of Al species. Pentahedral coordination ($Al^V$) were rarely on alumina and silica-alumina and reported to be a transition state to octacoordinated aluminum species and generated during calcination of γ-alumina.[1] Recently, it has been reported that $Al^V$ species on γ-alumina are surface active sites for stabilizing metal centers or nanoparticles to suppress sintering.[2] The $Al^V$–metal (e.g. Ba, Cu, Ru, Au, Ag, Pt and Pd) interaction was proposed to improve the catalytic activity of metal centers in various reactions,[3-6] such as $CO_2$ reduction and deNO(x) reactions. Therefore, $Al^V$ species have recently attracted great attention, particularly, they were proposed to be coordinatively unsaturated surface centers of supports for anchoring noble metal atoms for emerging single-atom catalysts.[2,7]

However, the previous reports showed that $Al^V$ species are not highly populating the surface. Although a certain amount of $Al^V$ species has been found to be generated during phase transformation from γ-$Al_2O_3$ to α-$Al_2O_3$ (up to 17 %),[1,8,9] only a very small amount of $Al^V$ species (< 2 at.%) was stabilized on the surface of γ-$Al_2O_3$, when suitable hydroxides or oxides such as $La_2O_3$ and BaO were added to inhibit the phase transformation and stabilize the unsaturated Al ions. It was also reported that $Al^V$ species only exist nearby $Al^{VI}$ species on the surface of crystalline $Al_2O_3$ and might represent $Al^{VI}$ in the vicinity of an oxygen vacancy, such as the defect spinel structures of the transition aluminas.[10] For silica-aluminas, $Al^V$ species were proposed to be located on the interface between alumina and silica or alumina and aluminosilicates.[11,12] Therefore, $Al^V$ species on $Al_2O_3$ or corresponding interfaces were assumed to be poorly distributed on the surface and hardly available for the reactants, which reinforced the doubt that $Al^V$ species are promising active centers for building high-performance catalysts such as emerging solid acids, single-atom catalysts and spatially confined catalysts due to their poor accessibility.[2,7]

In this study, we discovered a new type of $Al^V$ species in amorphous silica-alumina (ASA). It is homogeneously distributed and highly populates the surface, thus providing high accessibility for guest molecules. The ASA and $Al_2O_3$ nano-particles were prepared by flame spray pyrolysis as described previously, which could offer strong acidity.[13] All prepared particles had a size around 5-10 nm and their crystalline or amorphous structure has been verified by XRD. As shown in Fig. S1, both ASA samples, containing 10 and 30 at.% of aluminum (SA/10 and SA/30), only showed a broad reflection at 22-23° due to amorphous silica. As a reference, a pure $Al_2O_3$ sample SA/100 obtained without adding silica during synthesis has been investigated by XRD, as also shown in Fig. 1. Broad and weak signals of crystalline $Al_2O_3$ were observed for SA/100 due to its small particle size. The very fine nanoparticles with well-ordered lattice structure did not show significant diffraction and could therefore not be identified by XRD. High-resolution transmission electron microscopy (HRTEM) has also been applied to examine the existence of alumina phase domains in ASAs.[14-16] For SA/100, the image clearly revealed the well-ordered alumina lattice in Fig. 2a. For ASA samples, the alumina lattice disappeared, and the amorphous nature of both samples was corroborated, as shown in the HRTEM images in Fig. 2b and c. No small alumina clusters were detected on the ASA surface. As revealed by EDX atom mapping images, Al species were homogeneously distributed in the silica network of SA/10 (Fig. 2d-e) and SA/30 (Fig. 2f and g), again, no aggregated aluminum species or small alumina nanoparticles were observed on the surface.

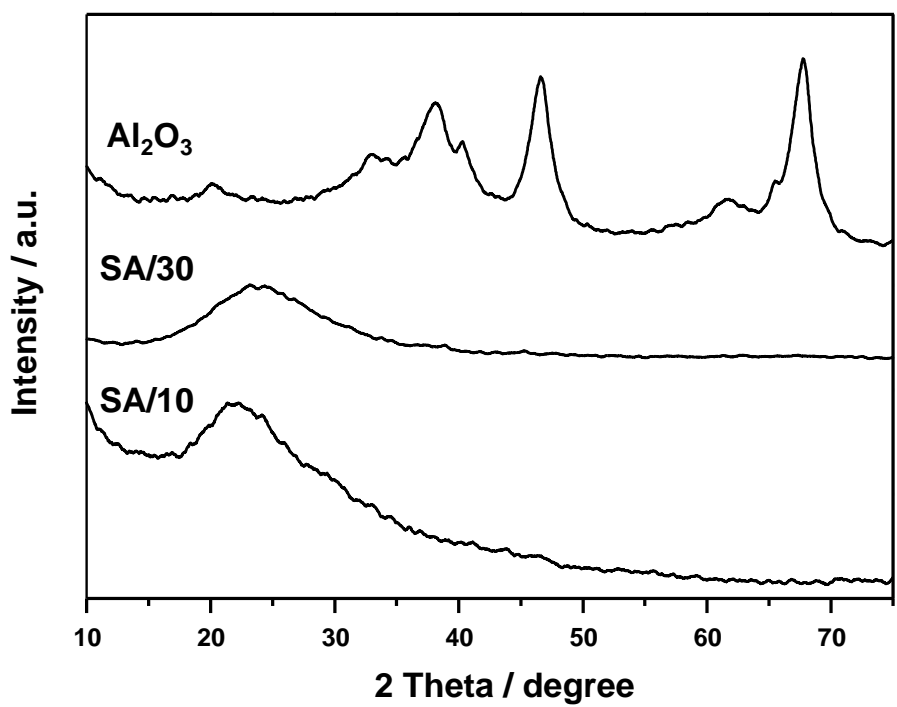

**Figure 1.** XRD pattern of SA/10, SA/30 and $Al_2O_3$.

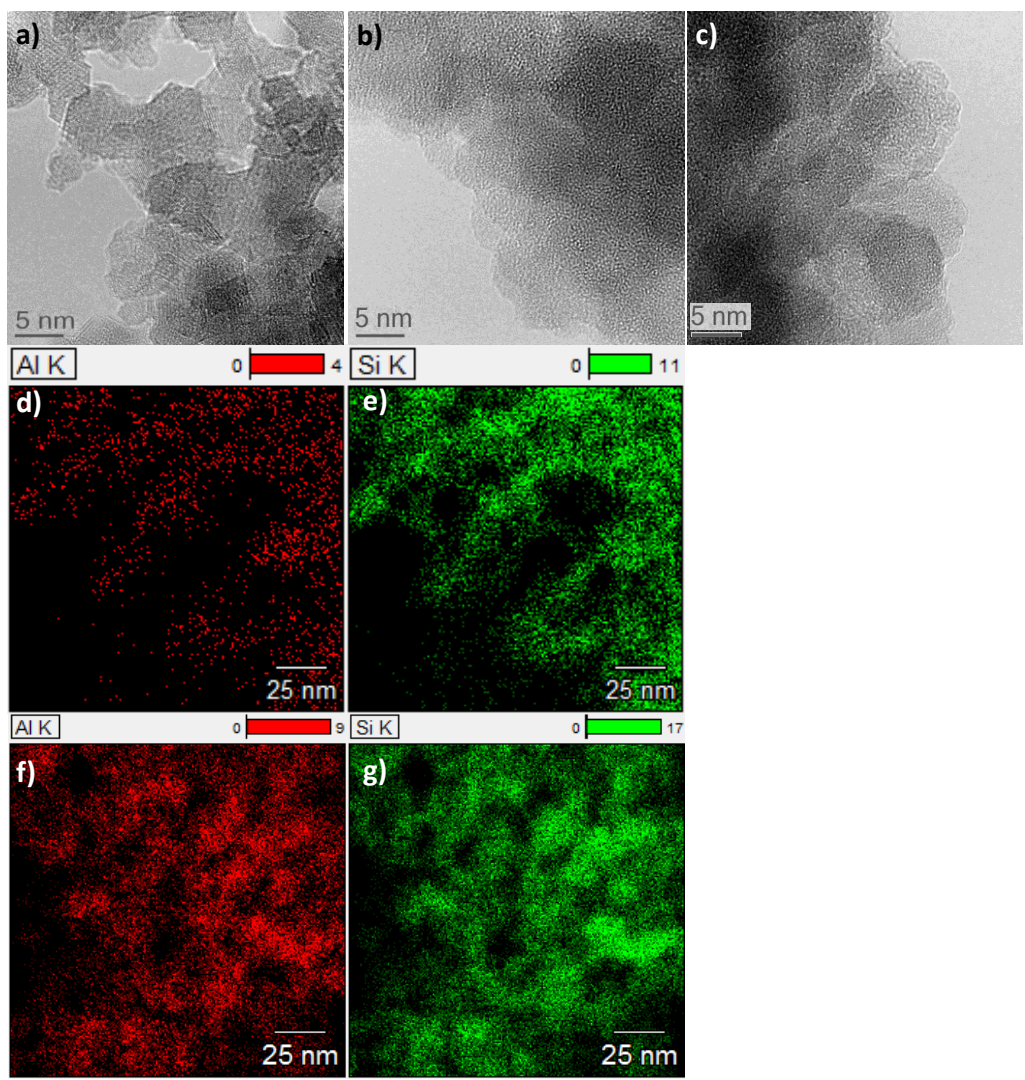

**Figure 2.** HRTEM images of a) SA/100, b) SA/10 and c) SA/30 and EDX images of SA/10 d) and e) and SA/10 f) and g).

To identify the local coordination of these homogenously dispersed Al species, solid-state $^{27}$Al MAS NMR spectroscopy combined with multiple quantum MAS (MQMAS) has been used to yield the necessary resolution for discriminating among nuclear quadrupole-interaction-broadened signals of the different Al species.[17-19] For both the SA/10 and SA/30 samples, Al$^V$ species have been clearly identified by $^{27}$Al MQMAS NMR spectroscopy. The subscript "de" and "hy" are dedicated to species present in dehydrated and hydrated state, respectively. As shown in Fig. 3a and 3c, two strong signals at (50, 55) and (23, 29) were assigned to Al$^{IV}$ and Al$^V$ species, respectively, indicating their predominant population in the aluminate species on both dehydrated SA/10 and SA/30 samples. Interestingly, only a small peak at (-9, 0.5) and a very weak signal at (-5.4, 2.7) were observed indicating the existence of a very small amount of $Al^{VI}$ species on the dehydrated SA/10 and SA/30 samples, respectively. This observation is distinct from the previous report where it was proposed that $Al^V$ species only exist nearby the alumina phase containing predominantly $Al^{VI}$ species.[10]

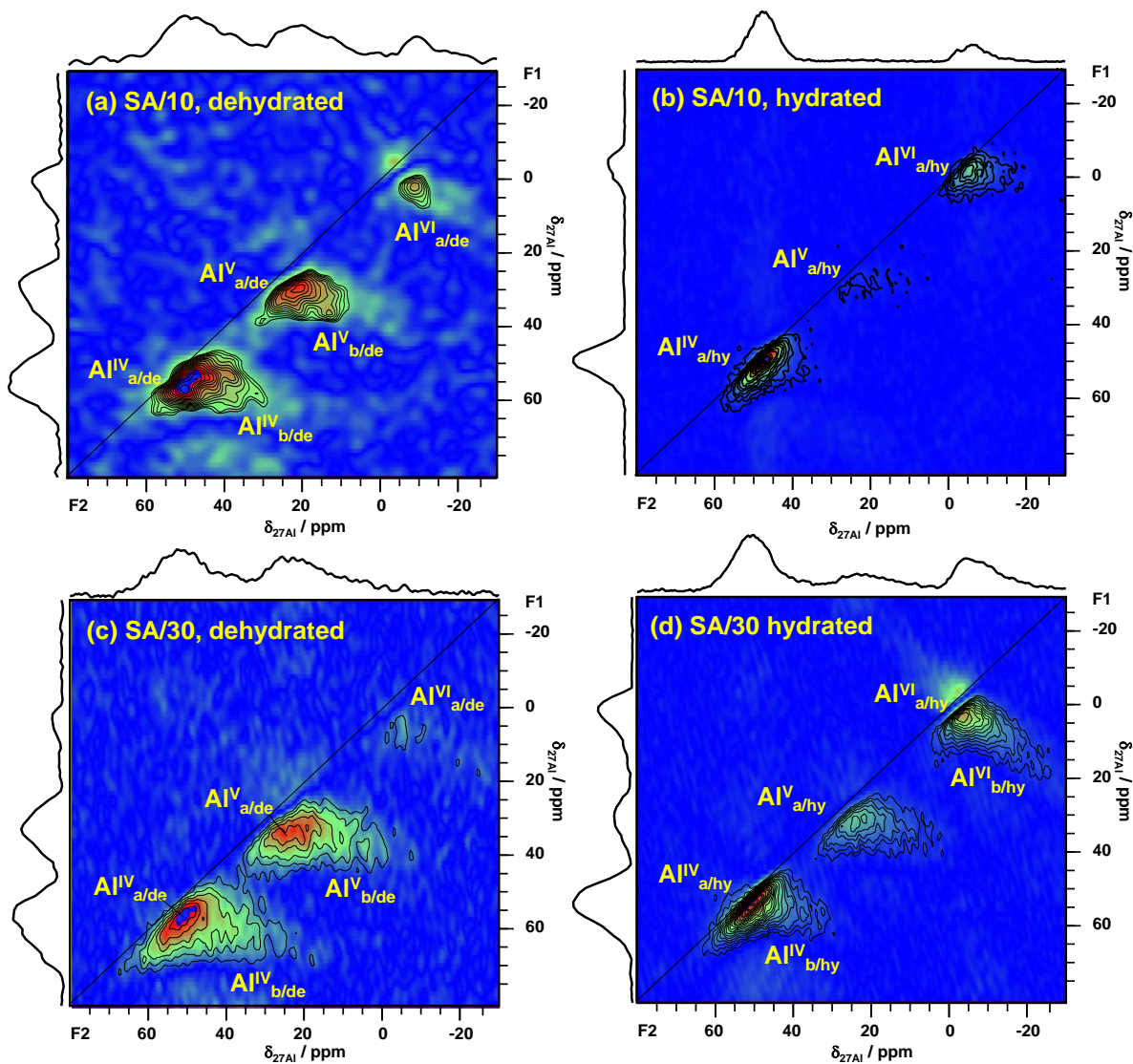

**Figure 3.** $^{27}Al$ MQMAS NMR spectra of dehydrated SA/10 (a) and SA/30 (b) and in corresponding rehydrated state (c) and (d).

To quantify the mole fraction of $Al^V$ species in ASA, 1D $^{27}Al$ MAS NMR spectra of dehydrated and hydrated SA/10 and SA/30 (shown in Fig. 4) were acquired at high field of 16.4 T and high MAS spinning of 14 kHz. The decomposition and quantitative evaluation of each signal were based on the parameters obtained from $^{27}Al$ MQMAS NMR spectra (isotropic chemical shifts, quadrupole coupling constant, and asymmetry parameters listed in Table 1.

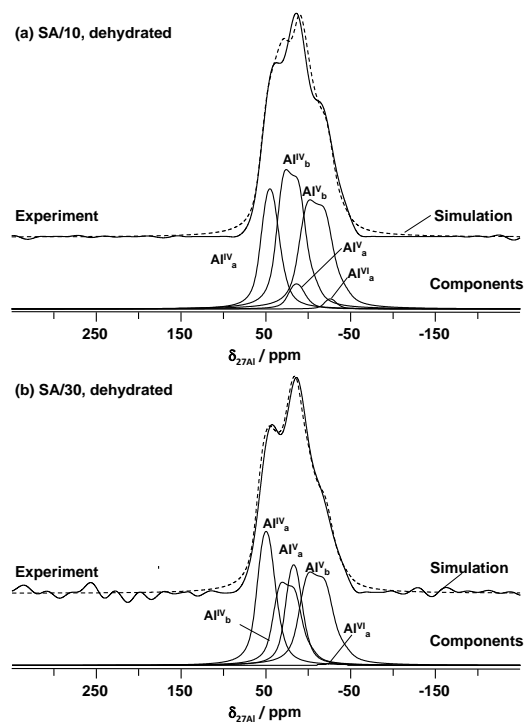

**Figure 4.** Deconvolution spectra of $^{27}$Al MAS NMR spectra (obtained at 700 MHz magnetic fields) of dehydrated SA/10 (a) and SA/30 (b), using the parameters listed in Table 1.

**Table 1.** Isotropic chemical shifts ($\delta_{iso}$), quadrupole coupling constant ($C_{QCC}$) and asymmetry parameters of the electric field gradient tensor ($\eta$) of each aluminum species, determined by simulation of $^{27}$Al MAS NMR spectra of dehydrated and rehydrated SA/10 and SA/30.

|  | SA10 | | | | | | SA30 | | | | | |
|---|---|---|---|---|---|---|---|---|---|---|---|---|
|  | dehydrated | | | hydrated | | | dehydrated | | | hydrated | | |
|  | $\delta_{iso}$ ppm | $C_{QCC}$ MHz | $\eta$ | $\delta_{iso}$ ppm | $C_{QCC}$ MHz | $\eta$ | $\delta_{iso}$ ppm | $C_{QCC}$ MHz | $\eta$ | $\delta_{iso}$ ppm | $C_{QCC}$ MHz | $\eta$ |
| Al$^{IV}_a$ | 58 | 7.7 | 0.5 | 58 | 7.7 | 0.5 | 58 | 7.7 | 0.5 | 58 | 7.7 | 0.5 |
| Al$^{IV}_b$ | 45 | 11.1 | 0.3 |  |  |  | 47 | 11.1 | 0.3 | 47 | 9.6 | 0.5 |
| Al$^{V}_a$ | 30 | 7.4 | 0.5 | 30 | 7.4 | 0.5 | 31 | 7.5 | 0.5 | 31 | 7.5 | 0.5 |
| Al$^{V}_b$ | 22 | 10.9 | 0.3 |  |  |  | 20 | 10.9 | 0.3 |  |  |  |
| Al$^{VI}_a$ | -5.6 | 6.3 | 0.8 | 1 | 5.1 | 0.5 | -3 | 7.6 | 0.8 | 1 | 5.1 | 0.5 |
| Al$^{VI}_b$ |  |  |  |  |  |  |  |  |  | 5 | 8.6 | 0.3 |

The obtained mole fractions of each species had been summarized in Table 2. The mole fraction of $Al^V$ species was 36.9 % in dehydrated SA/10 (with 61.9 % $Al^{IV}$ species and 1.2 % $Al^{VI}$ species) and 48.7 % in SA/30 (with 51.2 % $Al^{IV}$ species and 0.1 % $Al^{VI}$ species), respectively. The virtual absence of the $Al^{VI}$ species in both samples corroborated the findings of the TEM and EDX measurements that no alumina phase was existing in the silica-alumina materials. As previously reported, $Al^V$ species are generated only in the alumina phase or nearby the alumina phase with dominant $Al^{VI}$ species as encountered in phase transformation of $Al_2O_3$,[1,8] in the defect spinel structures of the transition aluminas,[20,21] or in the interface between alumina and silica or silica-alumina.[11,12]

**Table 2.** Concentration of each aluminum species (mol.%), determined by simulation of $^{27}Al$ MAS NMR spectra of dehydrated SA/10 and SA/30.

|  | $Al^{IV}_a$ % | $Al^{IV}_b$ % | $Al^V_a$ % | $Al^V_b$ % | $Al^{VI}_a$ % |
|---|---|---|---|---|---|
| SA/10, dehydrated | 24.7 | 37.2 | 4.4 | 32.5 | 1.2 |
| SA/30, dehydrated | 28.7 | 22.5 | 18.5 | 30.2 | 0.1 |

Here, we discovered that a large amount of $Al^V$ species can be produced (mole fraction nearly 50%) on amorphous silica-alumina without concomitant formation of an alumina phase or cluster. As commonly achieved in the synthesis of the aluminosilicate acids, minimizing the formation of the alumina phase or $Al^{VI}$ species during the synthesis can maximize the dispersion of Al atoms and the generation of $Al^{IV}$ species in the silica network or framework for optimal acidity. Therefore, our discovery of the formation of $Al^V$ species independent of the presence of an alumina phase is significant to overcome the current challenge of synthesizing highly dispersed $Al^V$ species for high performance catalysts or active materials such as strong acids and emerging single atom catalyst.

Interestingly, both $Al^V$ and $Al^{IV}$ species could co-exist in ASA with high population (mole fraction of nearly 50% for each species). It was assumed that $Al^{IV}$ species are available on the surface of ASA or zeolites, but $Al^V$ species were considered not to exist in high concentration on the surface. The surface $Al^{IV}$ species in silica networks or frameworks can interact with adsorbed water molecules to be transformed to $Al^{VI}$ species via partial hydrolysis.[22-25] To investigate whether $Al^V$ species are highly dispersed on the surface and accessible for guest molecules, the dehydrated SA/10 and SA/30 samples were exposed to water molecules and further investigated using $^{27}Al$ MQMAS NMR spectroscopy. As shown in Fig. 3b and 3d, the signal for Al$^V$ species was dramatically reduced in 2D NMR spectra and the signals of both Al$^{IV}$ and Al$^{VI}$ species were enhanced. Obviously, most of the Al$^V$ species, formed independently from the alumina phase, were accessible surface species, which are active and interact with adsorbed water molecules. They are mainly transferred to Al$^{VI}$ species via partial hydrolysis. Much more Al$^V$ than Al$^{IV}$ species were involved in the hydrolysis on the surface, which indicates that Al$^V$ species are more accessible to guest molecules and more effective in driving reactions. Al$^V$ species are found to be preferentially located on the surface. After introducing a small amount of aluminum (10%), nearly all Al$^V$ species were located on the surface of SA/10 and disappeared after hydration. When increasing the aluminum content up to 30%, part of the Al$^V$ species were formed inside the bulk of the particles, and thus, were not available for hydrolysis.

Thus, Al$^V$ species are proposed to be homogeneously dispersed in the amorphous silica network via Al$^V$–O–Si on the surface and their formation is found to be independent of the existence of an alumina phase. To confirm this conclusion and gain further insights into the local structure and geometry of the alumina species, density functional theory (DFT) calculations were applied to find candidate structures for Al$^V$ species in both the dehydrated and the hydrated ASA samples as shown in Fig. 5. The possible structures were optimized at B3LYP/6-31g(d) theoretical level, and the $^{27}$Al chemical shifts and quadrupole parameters (Table S2) were predicted at B3LYP/6-311+G(d,p) theoretical level on the basis of optimized structures. Fig. 5a and c demonstrate that Al$^V$ species (as shown in Fig. 5) could be stabilized and located in the vicinity of SiOH groups on the surface. The corresponding theoretical $^{27}$Al chemical shifts are 26 and 21 ppm for Al$^V$ species, similar to the experimental data (20-31 ppm). Upon water adsorption (Fig. 5d and f), most of these Al$^V$ species should be converted into Al$^{VI}$ species at -5 ppm (Table 3) with two coordinated water molecules (Fig. 5d), while a fraction of them would be converted into Al$^{IV}$ species at 52 ppm (Fig. 5b and Table 3). These findings are in line with the NMR experimental results indicating strong decrease of Al$^V$ signals with enhanced intensities for Al$^{VI}$ and Al$^{IV}$ signals after adsorption of water. Moreover, these mobile water molecules are able to improve the symmetry of Al$^V$ species when coordinated, in good agreement with the lower $C_{QCC}$ values obtained with rehydrated samples. The $C_{QCC}$ values for surface available Al$^V$ species on three DFT predicted structures under dehydration were 12.1, 16.2, and 13.2 MHz, which changed to 7.7, 8.3, and 7.4 MHz after hydration, respectively. This behavior is in line with previous theoretical and experimental work on γ-Al$_2$O$_3$, silica-alumina, and zeolites, where high $C_{QCC}$ values were always found for Al species on the dehydrated samples and lower ones for highly hydrated surfaces.[16] Besides the aforementioned possibility, the Al$^V$ species could be formed through dehydration reaction between the hydroxyl groups, and thus Al$^V$ species could incorporate into the silica network without being disturbed by SiOH groups as shown in Fig. 5e. These Al$^V$ species can be transferred into Al$^{VI}$ species upon water loading as well as shown in Fig. 5f.

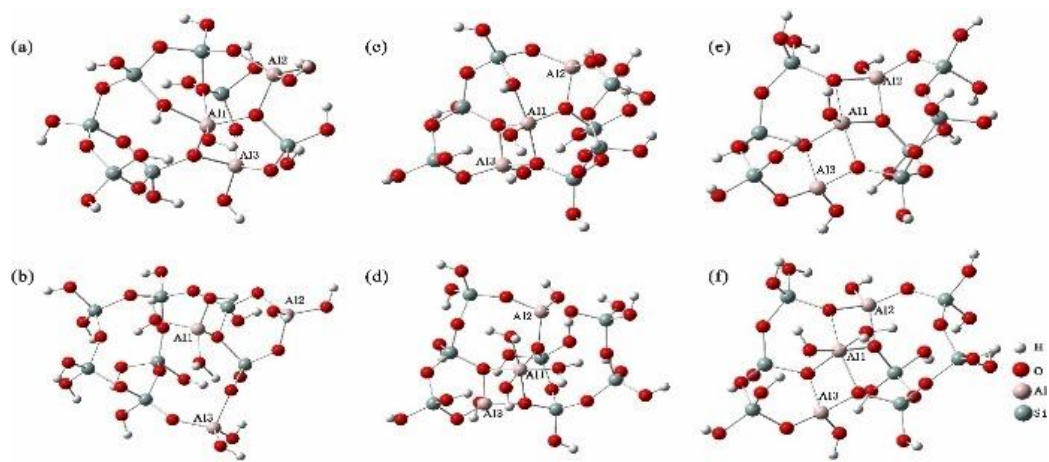

**Figure 4.** Optimized structure of Al$^V$ species in both dehydrated (a, c and e) and corresponding rehydrated (b, d and f) states calculated at B3LYP/6-31g(d) theoretical level.

**Table 3.** Isotropic chemical shifts ($\delta_{iso}$), quadrupole coupling constants ($C_{QCC}$) and asymmetry parameters of the electric field gradient tensor ($\eta$) of Al$^V$ species before (a, c, e) and after rehydration (b, d, f), determined from DFT calculations.

|     |                      | a    | b    | c    | d    | e    | f    |
|-----|----------------------|------|------|------|------|------|------|
| Al1 | $\delta_{iso}$ / ppm | 26   | 52   | 21   | -5   | 24   | -5   |
|     | $C_{QCC}$ / MHz      | 12.1 | 7.7  | 16.2 | 8.3  | 13.2 | 7.4  |
|     | $\eta$               | 0.79 | 0.98 | 0.31 | 0.65 | 0.41 | 0.27 |
| Al2 | $\delta_{iso}$ / ppm | 59   | 60   | 50   | 44   | 50   | 48   |
|     | $C_{QCC}$ / MHz      | 15.1 | 13.7 | 14.3 | 12.4 | 14.9 | 14.3 |
|     | $\eta$               | 0.66 | 0.79 | 0.53 | 0.67 | 0.77 | 0.55 |
| Al3 | $\delta_{iso}$ / ppm | 51   | 42   | 49   | 54   | 52   | 55   |
|     | $C_{QCC}$ / MHz      | 16.7 | 15.6 | 16.6 | 15.0 | 16.0 | 14.1 |
|     | $\eta$               | 0.27 | 0.31 | 0.92 | 0.78 | 0.88 | 0.90 |

With these findings, we conclude that, being structurally similar as $Al^{IV}$ species, $Al^V$ species can be formed with high population density and homogenous dispersion on the surface of silica-alumina. Unlike the $Al^V$ species generated in the alumina phase or interface of alumina with limited amount and surface availability, the new structural environment of $Al^V$ species discovered here promotes a high population of $Al^V$ species on the surface and these species are highly accessible for guest molecules. This finding could be of significance in the strive to maximize the efficiency of Al species as unique active centers for new solid acids or single-atom catalyst.

**Experiment Section**

HRTEM studies were carried out using a FEI Titan 80/300 ST unit. The instrument was equipped with a Cs corrector for spherical aberration of the objective lens. For imaging, a high-angle annular dark field (HAADF) detector, a GATAN post-column imaging filter was used. In all investigations, the microscope was operated at an acceleration voltage of 300 kV.

$^{27}$Al MQMAS NMR spectra were recorded on a Bruker Avance III 700 spectrometer with a 16.4 Tesla superconducting magnet operating at a resonance frequency of 182.5 MHz and 700 MHz for $^{27}$Al and $^1$H nuclei respectively. The samples were packed into 4 mm zirconia MAS rotors fitted with a Kel-F® cap inside a glove box to prevent any exposure to ambient air and moisture. The rotors were spun to 14 kHz MAS in a with 4 mm H-X double resonance probe. The MQMAS spectra were acquired using the three-pulse z-filter MQMAS pulse sequence[26] with pulse lengths of 5.5, 1.6, and 20 μs, a repetition time of 200 ms, and 64 $t_1$ increments. The corresponding 1D $^{27}$Al MAS NMR spectra were measured on the same spectrometer after single-pulse π/6 excitation with repetition times of 10 s to ensure complete relaxation of all the signals.

DFT calculations were carried out with the Gaussian09 program.[27] During the calculations, the structures of dehydrated and hydrated Al species were fully optimized using the B3LYP (Becke, three-parameter, Lee-Yang-Parr) method with 6-31G(d) basis set. The gauge-independent atomic orbital (GIAO) formalism[28] was adopted to predict $^{27}$Al shielding parameters at the B3LYP/6-311+G(d,p) level on the basis of the optimized structures. Taking the experimental $^{27}$Al chemical shifts (experimental result at 60 ppm, corresponding to the calculated absolute shielding at 495 ppm) of tetrahedral-coordinated Al in ZSM-5 sample as a benchmark[29], the calculated isotropic $^{27}$Al NMR chemical shifts were obtained.

**Corresponding author**

*jun.huang@sydney.edu.au; zhenganm@wipm.ac.cn

**Acknowledgements**

J.H., Z.W., Z.L., and C.Z. acknowledge the financial supports from Australian Research Council Discovery Projects (DP150103842) and the Faculty's MCR Scheme, Energy and Materials Clusters at the University of Sydney. A.Z. and F.D. acknowledge the support by the National Natural Science Foundation of China (21522310, 21473244 and 21210005). The Mark Wainwright Analytical Centre at UNSW is acknowledged for access to the solid state NMR spectrometer funded through ARC LIEF LE0989541.